\documentclass[universe,article,accept,pdftex,moreauthors]{Definitions/mdpi} 

\firstpage{1} 
\pubvolume{10}
\issuenum{120}
\articlenumber{439}
\pubyear{2024}
\copyrightyear{2024}
\datereceived{21 October 2024} 
\daterevised{25 November 2024} 
\dateaccepted{26 November 2024} 
\datepublished{28 November 2024} 
\hreflink{https://doi.org/10.3390/\linebreak universe10120439} 

\usepackage{aas_macros}
\usepackage{bm} 
\usepackage{bbm}
\usepackage[labelformat=simple]{subcaption}

\DeclareCaptionLabelFormat{subcaptionlabel}{\normalfont(\textbf{#2}\normalfont)}
\usepackage{siunitx}

\Title{
An Analysis of Variance of the Pantheon+ Dataset: Systematics in the Covariance Matrix?
}

\TitleCitation{%
An Analysis of Variance of the Pantheon+ Dataset: Systematics in the Covariance Matrix?
}

\newcommand{\mat}[1]{\ensuremath{\bm{\mathsf{#1}}}} 
\newcommand{\vect}[1]{\ensuremath{\bm{#1}}} 
\newcommand{\one}{\ensuremath{\mathbbm{1}}} 
\newcommand{\lcdm}{\ensuremath{\Lambda\text{CDM}}}
\renewcommand{\citep}{\cite}

\newcommand{\ccor}{\ensuremath{\widetilde{\mat{C}}}}

\Author{{Ryan~E.} 
 {Keeley} 
 $^{1
}$\orcidA{ }, 
Arman Shafieloo $^{2
}$\orcidB{ } and 
Benjamin L'Huillier $^{3,}$*
\orcidC{ }
}

\AuthorNames{Ryan E. Keeley, Arman Shafieloo and Benjamin L'Huillier}

\AuthorCitation{{Keeley, R.E.;} 
Shafieloo, A.; L'Huillier, B.}

\address{%
$^{1}$ \quad Department of Physics, University of California Merced, 5200 North Lake Road, Merced, CA 95343, USA; rkeeley@ucmerced.edu\\
$^2$ \quad Korea Astronomy and Space Science Institute,  776 Daedeok-daero, Yuseong-gu, Daejeon 34055, \mbox{Republic of Korea;} shafieloo@kasi.re.kr\\
$^{3}$ \quad Department of Physics and Astronomy, Sejong University, 05006 Seoul, Republic of Korea}

\corres{Correspondence: benjamin@sejong.ac.kr%
}

\abstract{%
We investigate the statistics of the available Pantheon+ dataset. Noticing that the $\chi^2$ value for the best-fit $\Lambda$CDM model to the real data is small, we quantify how significant its smallness is by calculating the distribution of $\chi^2$ values for the best-fit $\Lambda$CDM model fit to mock Pantheon+-like datasets, using the provided covariance matrix.
We further investigate the distribution of the residuals of the Pantheon+ dataset with respect to the best-fit $\Lambda$CDM model, and notice that they scatter less than would be expected from the covariance matrix but find no significant kurtosis.  These results  point to the conclusion that the Pantheon+ covariance matrix is over-estimated. One simple interpretation of these results is a $\sim$7\% overestimation of errors on SN distance moduli in Pantheon+ data. 
When the covariance matrix is reduced by subtracting an intrinsic scatter term from the diagonal terms of the covariance matrix, the best-fit $\chi^2$ for the $\Lambda$CDM model achieves 
a normal value of 1580 and no deviation from $\Lambda$CDM is detected. We further quantify how consistent the $\Lambda$CDM model is with respect to the 
modified data with the subtracted covariance matrix using model-independent reconstruction techniques such as the iterative smoothing method. We find that the standard model is consistent with the data. 
There are a number of potential explanations  for this smallness of the $\chi^2$, such as a Malmquist bias at high redshift, or  accounting for systematic uncertainties by adding them to the covariance matrix, thus approximating systematic uncertainties as statistical ones.
}


\begin{document}

\section{Introduction}


%
One of the key datasets that built the standard model of cosmology is the compilation of Type Ia supernovae (SN). 
{While,} 
 by the 1990s, increasing evidence for a cosmological constant emerged from the age of the Universe \cite{1995Natur.376..399B, 1996MNRAS.282..926J} to the large-scale structures \cite{1990Natur.348..705E, 1995Natur.377..600O},
SNIa gave the first direct evidence
that the expansion of the Universe was accelerating~\citep{SupernovaSearchTeam:1998fmf,SupernovaCosmologyProject:1998vns}.  
The most recent SN dataset is the Pantheon+ compilation~\citep{2022ApJ...938..110B,2022ApJ...938..113S}, which builds on the earlier Pantheon analysis~\citep{Pan-STARRS1:2017jku}.\endnote{After this article was submitted, the Union-3 compilation \cite{2023arXiv231112098R} was posted on the arXiv, although the data were not publicly available yet.}   

SN distances are inferred from SN light curves.  
It is assumed that Type Ia SN compose a standardizable candle, and thus, by measuring their light curves, one can then also model various systematic effects and robustly infer the distance modulus of each SN, up to an overall scaling~(see, e.g., \cite{2020ApJ...899....9K}). 
{ {The} stretch and color of the SN light curve are involved in the SN standardization process~\cite{1998A&A...331..815T} and contemporary analyses included additional corrections for selection effects and the host galaxy of the SN \cite{2022ApJ...938..110B}.
}
All of the statistical and systematic uncertainties are then combined into a single covariance matrix for the distance modulus of each individual SN.

The Pantheon+ dataset \citep{2022ApJ...938..110B,2022ApJ...938..113S}, along with its various systematic checks ~\citep{2021arXiv211003486B,2022ApJ...938..112P,2022PASA...39...46C,2022ApJ...938..111B,2021arXiv211204456P,2021ApJ...909...26B}, provides a covariance matrix, so the presumption is that all of the statistical information in the Pantheon+ data can be captured by two-point statistics and that the noise in the data should follow a multivariate Gaussian distribution.  We show that the residuals of Pantheon+ do not follow a multivariate Gaussian distribution, in that the $\chi^2$ value for the flat $\Lambda$CDM model that best-fits the data is too small to have arisen from a multivariate Gaussian characterized by the provided Pantheon+ covariance matrix.

It is important to test the  presence of systematics in cosmological datasets, since such systematics could be mistakenly interpreted as a cosmological effect (e.g., \cite{2019MNRAS.485.2783L}).

\section{Results\label{sec:Results}}

We first noticed that the $\Lambda$CDM model that best fits the Pantheon+ data yields a low value for 
$\chi^2=1387.10$
This is suspiciously small for a dataset with 1580 data points. In order to quantify this, we 
use the same data selection criteria
as recommended by the Pantheon+ team, namely that each kept SN must have 
$z>0.01$,
and not be used in the SH0ES calibration. {We use the $z_{\rm hd}$ redshifts that the Pantheon+ dataset provides.}  We calculate the distance modulus $\mu(z)$ for each SN, allowing for the standard $\Lambda$CDM parameters to vary $(H_0, \Omega_\mathrm{m})$, and also vary the nuisance parameter $M_B$, the absolute magnitude of Type Ia SN.  We first calculate the standard (flat) $\Lambda$CDM expansion history,
\begin{equation}
    H(z) = H_0 \sqrt{\Omega_\mathrm{m}(1+z)^3 + 1-\Omega_\mathrm{m}},
\end{equation}
and integrate it to calculate the luminosity distance,
\begin{equation}
    D_L(z) = (1+z) \int_0^z dz' \frac c { H(z')},
\end{equation}
which is used to calculate the distance modulus,
\begin{equation}
    \mu(z) = 5 \log_{10} \left(\frac{D_L(z)}{\SI1{Mpc}}\right) + 25. 
\end{equation}
{The} 
Pantheon+ data are uncalibrated and so $M_B$ needs to be {added to} the distance modulus to calculate the theoretical values for $m_B$ for each SN
\begin{equation}
    m^{\rm theory}_{B,i} = \mu(z_i) + M_B.
\end{equation}
{Thus,} the $\chi^2$ takes the following form
\begin{equation}
    \chi^2 = \sum_{i,j} (m^{\rm data}_{B,i} - m^{\rm theory}_{B,i})^{T}  \mathcal{C}_{i,j}^{-1} (m^{\rm data}_{B,j} - m^{\rm theory}_{B,j})
\end{equation}
where $i, j$ span the set of SN in the dataset and $\mathcal{C}$ is the Pantheon+ covariance matrix.

A $\chi^2$ distribution with 1580 degrees of freedom would predict a value of 
$\chi^2=1387.10$
with a probability of 
\num{1.3e-5}.
{{For} such a large number $N$, the $\chi^2$ distribution with $N$ degrees of freedom is approximately Gaussian with a mean $N$ and variance $\sqrt{2N}$, which makes the observed value a 3.4 $\sigma$ outlier.}
This is not a direct comparison since, with real data, one has to marginalize over the unknown parameters of the $\Lambda$CDM model.  To this end, we calculate the probability of finding such a small best-fit $\chi^2$ value by generating a set of 10,000 mock Pantheon+-like datasets generated from a flat $\Lambda$CDM model with $H_0 = \SI{70}{km.s^{-1}.Mpc^{-1}}$, $\Omega_\mathrm{m}=0.3$, and $M_b=-19.0$. Distance moduli evaluated from this model, evaluated at the redshifts of the Pantheon+ SNe, then have random noise added to them, as drawn from a multivariate Gaussian characterized by the Pantheon+ covariance matrix.   With {{each} of these} these datasets, we calculate the best-fit $\Lambda$CDM model.  The $\chi^2$ values for these best-fit $\Lambda$CDM models for these 10,000 datasets then form a distribution.  We can then evaluate the probability of observing a best-fit 
$\chi^2=1387.10$
by seeing where it lies in this distribution.

\begin{figure}[H]
    
    \includegraphics[width=\columnwidth]{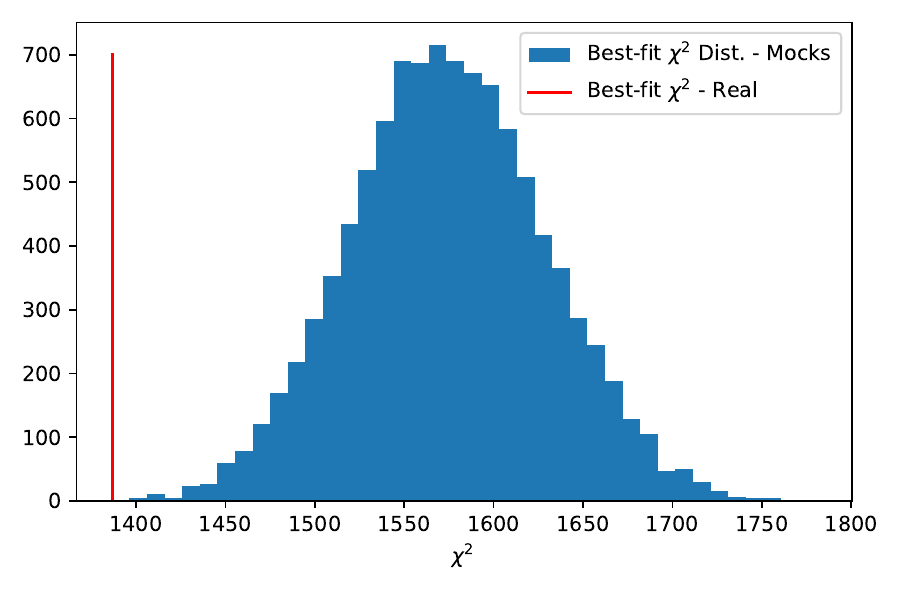}
    \caption{The distribution of $\chi^2$ values from a flat $\Lambda$CDM model that best fits mock data generated from a flat $\Lambda$CDM model with $H_0 = 70$ km/s/Mpc and $\Omega_\mathrm{m}=0.3$ and $M_b=-19.0$ (blue). The $\chi^2$ value for the best-fit flat $\Lambda$CDM model to the real data is shown in red.}
    \label{fig:chi2dist}
\end{figure}

The results can be seen in Figure~\ref{fig:chi2dist}, where {{none of}} our 10,000 realizations had \mbox{$\chi^2 < 1387.10$.}  This corresponds to a greater than $3.9$-$\sigma$ deviation, indicating the possible presence of an unaccounted for systematic.
The task now is to identify possible candidates for this possible systematic.

To this end, the next calculation we perform is visualizing the residuals with respect to the best-fit $\Lambda$CDM model of the real Pantheon+ data, normalized to the correlated uncertainties.  The idea is to show how often the residuals scatter with an amplitude of 1,2,3-$\sigma$, as measured by the covariance matrix. This is to test whether the residuals are distributed according to a multivariate normal distribution. The existence of non-zero off-diagonal components in the covariance matrix makes this computation slightly tricky. Since the individual datapoints are correlated, the variables that would independently statistically fluctuate around the mean are linear combinations of the datapoints. To find this linear combination, we find the basis where the covariance matrix is diagonal and the entries of the residual vector in that basis would be independent.

To show this calculation, we start with the definition of the $\chi^2$ statistic, which is invariant under basis transformations,
\begin{equation}
    \chi^2 = \vect X^\intercal \mat{C}^{-1} \vect X
\end{equation}
where $\vect X$ is the residual and $\mat{C}^{-1}$ is the inverse covariance matrix.  We diagonalize the inverse covariance matrix with its matrix of eigenvectors $\mat V$
\begin{equation}
    \chi^2 =\vect X^\intercal \mat V \mat V^{-1} \mat{C}^{-1} \mat V \mat V^{-1}\vect  X
\end{equation}
and so transforming the residual into the diagonal basis, 
\begin{equation}
    \tilde {\vect X} = \mat V^{-1} \vect X,
\end{equation}
we can define a normalized residual 
\begin{equation}
     \tilde X_{\rm{norm},i} = \tilde X_i w_i
\end{equation}
where $w_i$ is the square root of the eigenvalues of $\mat{C}^{-1}$, such that 
\begin{equation}
    \sum_i \tilde X_{\rm{norm},i}^2 = \chi^2.
\end{equation}
{Thus,} $\tilde X_{\rm{norm},i}$ is our normalized residual where each component is statistically independent.

\begin{figure}[H]
    
    \includegraphics[width=\columnwidth]{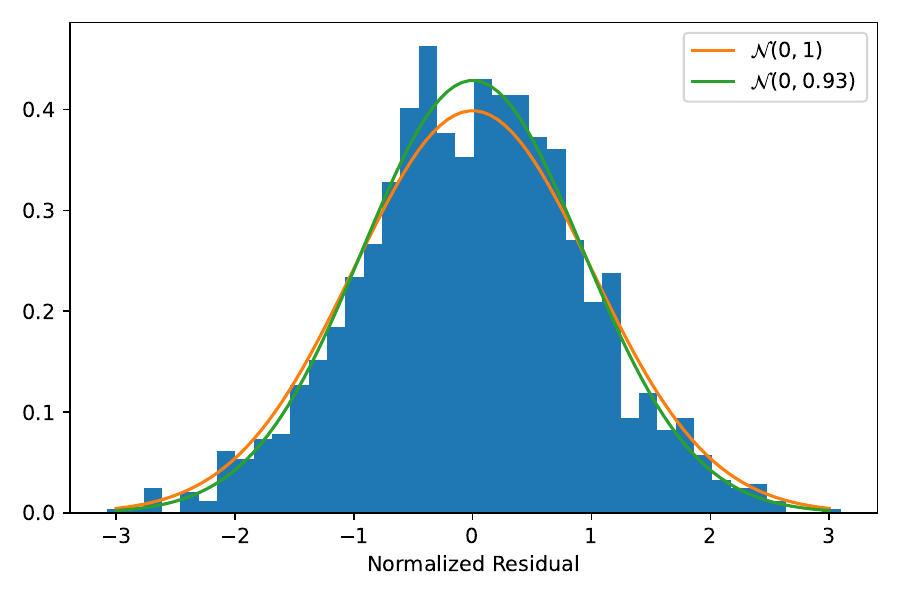}
    \caption{The distribution of the Pantheon+ residuals with respect to the best-fit $\Lambda$CDM model normalized to the correlated uncertainties in the Pantheon+ covariance matrix (blue). The residuals scatter less than would be expected from a standard normal distribution $\mathcal N(0,1)$ (orange).  {The green curve is a Gaussian distribution with a standard deviation of 0.93, the observed standard deviation of the normalized residuals of the Pantheon+ data}.}
    \label{fig:NormRes}
\end{figure}

We see the results of this calculation in Figure~\ref{fig:NormRes} where we see the normalized residuals scatter less than a standard normal distribution, as {{would} be} expected given the covariance matrix.  Specifically, the standard deviation of these normalized residuals is 0.93, indicating the errors on the distances inferred from the Pantheon+ SN are overestimated by 7\%{{, which} is just another manifestation} of the smallness of the $\chi^2$. 
{{In other} words}, we would find such a small standard deviation of the residuals at a frequency of less than 1 in 10,000 by random chance, the same frequency of the smallness of the $\chi^2$.

\begin{figure}[H]
    
    \includegraphics[width=\columnwidth]{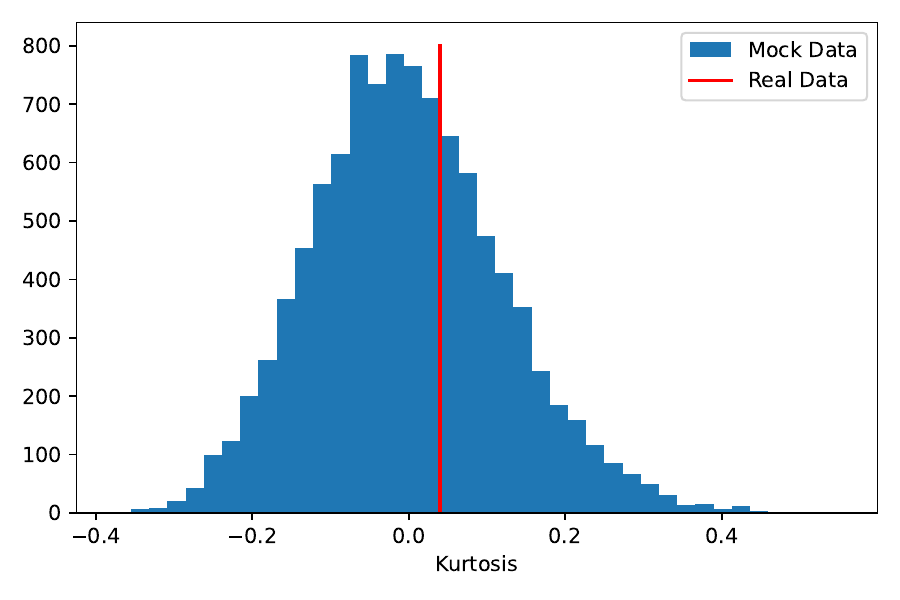}
    \caption{The distribution of kurtosis values for the normalized residuals of our mock Pantheon+-like datasets (blue). The kurtosis of real Pantheon+ normalized residuals are shown in red.}
    \label{fig:kurtosis}
\end{figure}

To check whether this feature {{arises}} from a non-zero kurtosis, we calculate the kurtosis of the Pantheon+ normalized residuals and calculate the significance of that number by calculating the kurtosis of the normalized residuals for our 10,000 mock Pantheon+-like datasets{{, as shown} in Figure~\ref{fig:kurtosis}}.  
The kurtosis for the real Pantheon+ data was found to be 0.04, which is well inside the distribution of the kurtosis for mock Pantheon+-like data.

On {{the}} one hand, we have a $\chi^2$ that is too small to explain, and at the same time, we have residuals that are Gaussian.
These points lead to the conclusion that the covariance of the Pantheon+ dataset is overestimated.

\begin{figure}[H]
    
    \includegraphics[width=0.495\columnwidth]{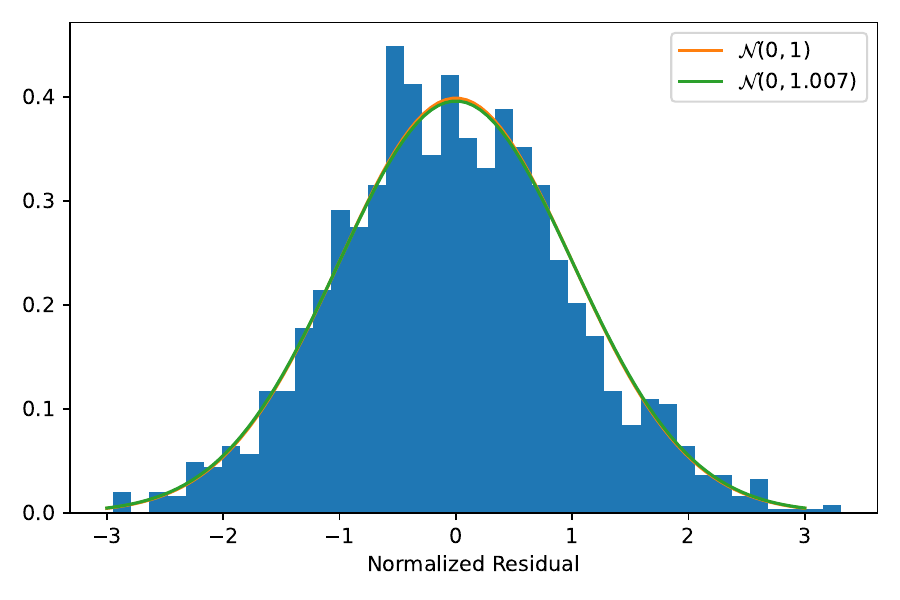}
    \includegraphics[width=0.495\columnwidth]{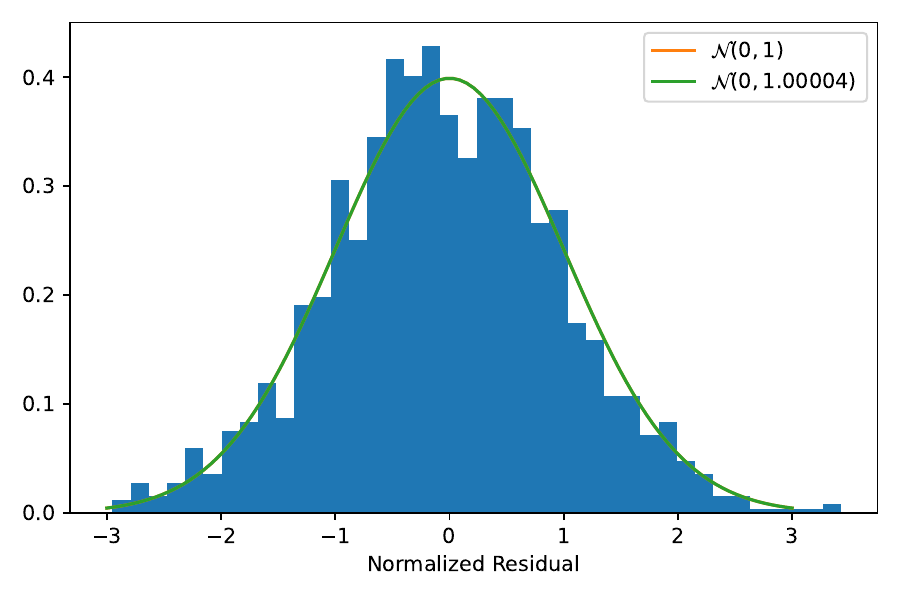}
    \caption{(\textbf{left}) {Normalized residuals for the case where the covariance matrix was scaled by the observed standard deviation of the normalized residuals of the Pantheon+ dataset, 0.93$^2$}. (\textbf{right}) Normalized residuals for the case where $\delta^2 = 0.002$ is subtracted from the covariance matrix. The green normal distribution in each of these panels has a standard deviation corresponding to the measured standard deviation of the normalized residuals for each case.}
    \label{fig:syn_corr}
\end{figure}

We investigate two cases that may have caused the covariance matrix to be overestimated.  The first is the case that  each element was scaled by the same value.  Since we find the standard deviation of the normalized residuals to be 0.93, we multiply the existing covariance matrix by 0.93$^2$ and perform the same calculation as previously.  We find that the best-fit $\chi^2$ is 1582.1 The second is the case {{of overestimated}} intrinsic scatter, which is \emph{{a priori} 
} unknown and modelled as a term added to each diagonal component of the covariance matrix. Such a term is supposed to model any unknown stochastic processes that would cause the brightness of individual SN to scatter beyond any measurement or calibration uncertainty. Thus, we subtract off $\delta^2=0.002$ from the diagonal of the covariance matrix.  That value was found such that the best-fit $\chi^2=1580$. We show the distribution of the normalized residuals for this case in Figure~\ref{fig:syn_corr}. {Subtracting an intrinsic scatter term from the covariance matrix or rescaling the covariance matrix will not offer an ultimate solution to the problem of the smallness of the $\chi^2$, but they are useful in quantifying the magnitude of the problem. Identifying an ultimate solution to the problem of the smallness of the $\chi^2$ is beyond the scope of this work since we do not have access to all of the ingredients that go into constructing the covariance matrix. }

\section{{Non-Parametric Reconstruction of the Expansion History}}

In order to assess whether this abnormally small $\chi^2$ arises from cosmology or systematics, we reconstruct the distance moduli--redshift relation using the iterative smoothing method, a model-independent method \citep{2006MNRAS.366.1081S,2007MNRAS.380.1573S}.
This data-driven method allows us to assess whether the problem may come from the assumed model or whether it is intrinsic to the data covariance matrix.  
We specifically use the iterative smoothing method in this work as it is useful for efficiently finding expansion histories that fit the data better than the best-fit $\Lambda$CDM model, and the amount of over fitting that results can be calibrated.

We applied the iterative smoothing method to 
the generated data points, {{taking} into account the} {(uncorrected)} covariance matrix \mat C from Pantheon+, to reconstruct the distance modulus.
Starting with an initial guess $\hat\mu_0$, we iteratively calculated
the reconstructed $\hat\mu_{n+1}$ at iteration $n+1$ as \cite{2018PhRvD..98h3526S}:
\begin{align}
  \hat\mu_{n+1}(z) & = \hat\mu_n(z) + \frac{\vect{\delta\mu}_n^\intercal \cdot  \mat C^{-1} \cdot \vect{W}(z)}
  { \one^\intercal \cdot \mat{C}^{-1} \cdot \vect{W}(z)},
  \intertext{where }
  \vect{W}_i(z) & =  {\exp{\left(- \frac{\ln^2\left(\frac
      {1+z}{1+z_i}\right)}{2\Delta^2}\right)}}
  \intertext{is a vector of weights, }
  \one & = (1,\dots,1)^\intercal\\
  \intertext{is a column vector, and}
  \vect{\delta\mu}_n|_i & = \mu_i-\hat\mu_n(z_i).
\end{align}
{We define} the $\chi^2$ value of the reconstruction $\hat{y}_n(t)$ as
\begin{equation}\label{eqn:chi2}
\chi_n^2 = \vect{\delta y}_n^\intercal \cdot \mat{C}^{-1} \cdot \vect{\delta y}_n.
\end{equation}

\begin{figure}[H]
     {\captionsetup{justification=centering}
     \begin{subfigure}[t]{0.495\textwidth}
         
         \includegraphics[width=\textwidth]{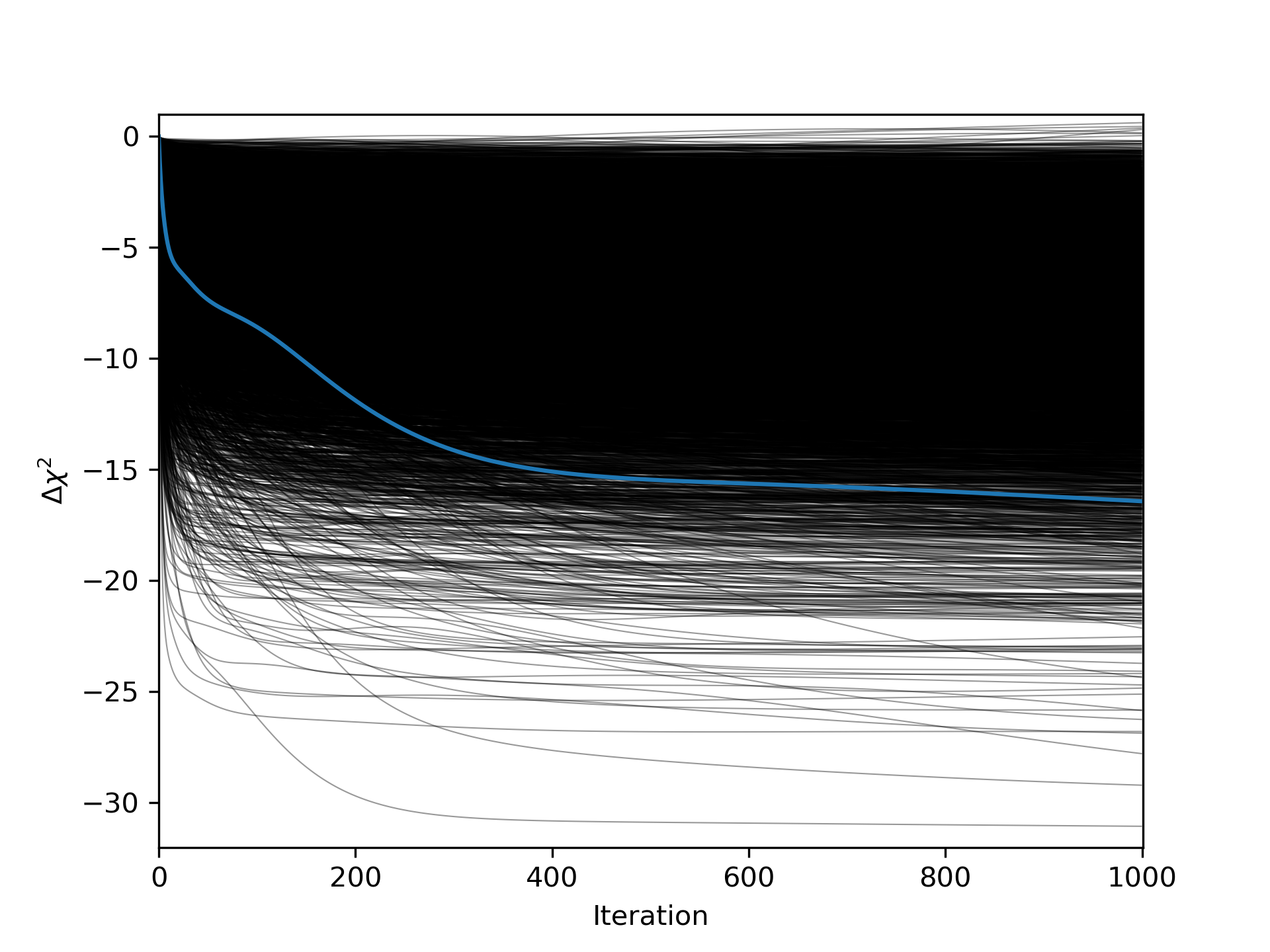}
         \caption{
         }
         \label{fig:dchi2}
     \end{subfigure}
     \hfill
     \begin{subfigure}[t]{0.495\textwidth}
         
         \includegraphics[width=\textwidth]{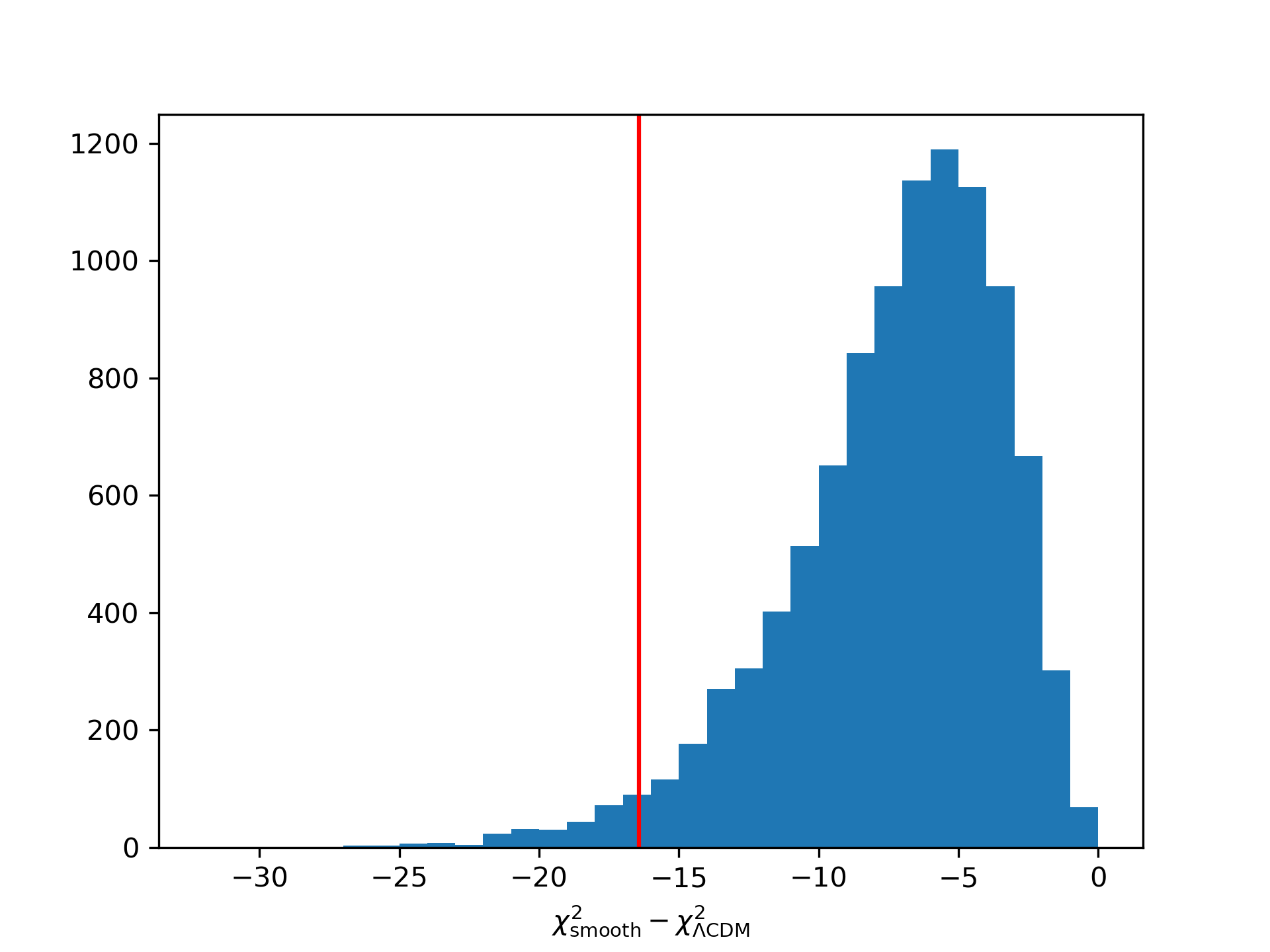}
         \caption{}
         \label{fig:dchi2Dist}
     \end{subfigure}}
        \caption{{Results} 
 for the iterative smoothing for the 10,000 mock realizations of the data using the corrected covariance matrix. (\textbf{a}) $\Delta\chi^2$ as a function of the iteration for our 10,000 mock {{datasets}}; \mbox{(\textbf{b}) The} distribution of the difference between the $\chi^2$ value that results from the iterative smoothing method {{at the} final iteration $N_\text{iter}=1000$} and the $\chi^2$ value of the best-fit $\Lambda$CDM model for our \mbox{10,000 mock} {{datasets}} is shown in blue. The difference between the $\chi^2$ value that results from smoothing and the $\chi^2$ value of the best-fit $\Lambda$CDM model for the real data is shown in red.}
        \label{fig:smoothing}
\end{figure}

Figure~\ref{fig:smoothing}a shows the evolution of $\Delta\chi^2 = \chi^2_n-\chi^2_0$, where $\chi^2_0$
 is the initial $\chi^2$, i.e., the $\chi^2$ of the \lcdm\ best fit, for the mock realizations (in black), and for the actual Pantheon+ data in blue. 
The Pantheon $\Delta\chi^2$ lies within the {{mocks}}, and there do not seem to be any anomalies. 
Figure~\ref{fig:smoothing}b shows the histogram of the $\Delta\chi^2${ {at the final} iteration $N_\text{iter}=1000$}. 
The $\chi^2$ value that results from the iterative smoothing method applied to the real Pantheon+ data lies within the distribution {{of}} $\Delta\chi^2$ values.
Thus, the small improvement in $\chi^2$ from the iterative smoothing method compared to $\Lambda$CDM is not significant and the iterative smoothing approach does not find any anomalies in the $\Delta\chi^2 $ of the Pantheon+ data. 

\vspace{-12pt}

We also calculate all the cases in Section~\ref{sec:Results}, 
but instead of the best-fit $\Lambda$CDM {{model}}, we use the 
{{final} iteration} of the iterative smoothing method.  
{{In other} words}, rather than calculate where the $\chi^2$ value of the best-fit $\Lambda$CDM model lies in the distribution of $\chi^2$ values of the best-fit $\Lambda$CDM model for mock data, we calculate where the $\chi^2$ value that results from the iterative smoothing method lies in the distribution of $\chi^2$ values that result from the iterative smoothing method for mock data.  
{{We} then also} calculate the distribution of residuals with respect to the results of the iterative smoothing method normalized to the correlated uncertainties.  
Since the result of the iterative smoothing method is so close to the $\Lambda$CDM model, the results for each of the cases in Figures~\ref{fig:chi2dist}--\ref{fig:syn_corr} are not changed significantly{{, regardless} of} whether the residuals are with respect to $\Lambda$CDM or the smoothing result. 
{The fact that the  smallness of the $\chi^2$ remains in the model-independent reconstruction case, i.e., it does not depend on the model assumption, points towards a problem in the covariance matrix rather than a cosmological origin.
It is important to note that the fact that the results of smoothing using the ``corrected covariance matrix'' are consistent with the \lcdm\ results suggests that no deviation from \lcdm\ was hiding in the ``uncorrected covariance matrix''. 
}

%

%

\begin{figure}[H]
    
    \includegraphics[width=\columnwidth]{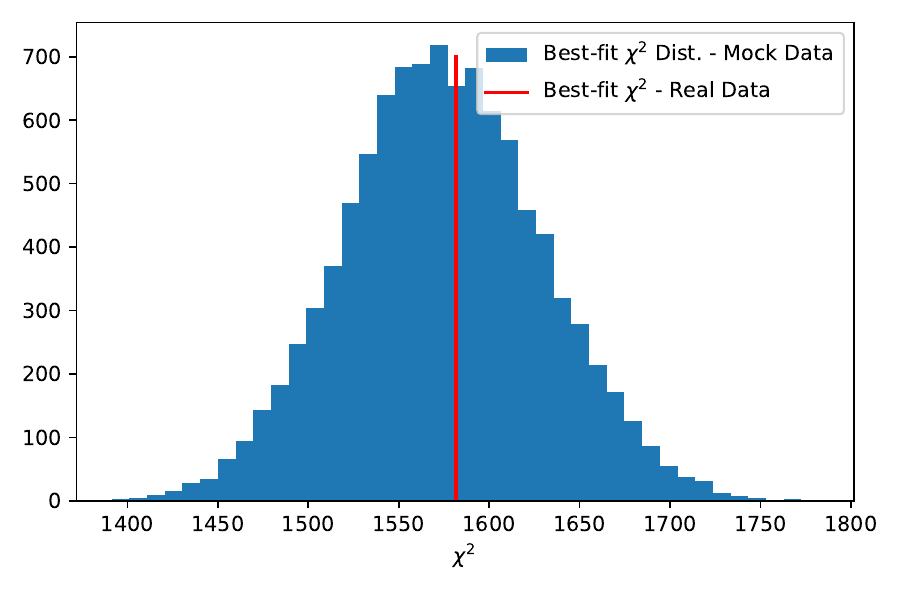}
    \caption{
    The distribution of best-fit $\chi^2$ values for mock data (blue) and for the real data (red){ {using} the corrected covariance matrix \ccor}.  For both, the best-fit $\chi^2$ values were calculated with the iterative smoothing method using the subtracted covariance matrix.  To compare like with like, the mock data were also generated with the subtracted covariance matrix, even though the distributions for the real covariance matrix and subtracted one are statistically consistent.     }
    \label{fig:chi2dist_smallcov}
\end{figure}

To show that subtracting an intrinsic scatter term from the covariance matrix makes everything consistent, we calculate a new distribution of $\chi^2$ values from mock data generated from this new subtracted covariance matrix {$\ccor = \mat C - \delta^2 \mat I$}.  The best-fit $\chi^2$ values for this case were calculated using the iterative smoothing method, both for the mock data and for the real data. The results of this case are shown in Figure~\ref{fig:chi2dist_smallcov}. The first feature to notice is that the distribution of $\chi^2$ values for this subtracted and smoothed case is not noticeably different from the case in Figure~\ref{fig:chi2dist}, where the data are generated from the original covariance matrix and fit with $\Lambda$CDM.  Thus, the distribution of best-fit $\chi^2$ values is determined less by the values of the elements of the covariance matrix but {{more}} by the fact that there are 1580 independent variables (combinations of data points).  The fact that the red line in Figure~\ref{fig:chi2dist_smallcov} sits at just under 1580 indicates that the value of $\delta^2=0.002$ that we subtract from the diagonal of the covariance matrix is robust to the choice of whether we fit the real data with the $\Lambda$CDM model or reconstruct an expansion history with the iterative smoothing method. This value of $\delta^2=0.002$ corresponds to subtracting off $\sim$27\% of the average error budget for each supernova. These results in turn indicate that there is no feature in the residuals of the real data that cannot be explained by the $\Lambda$CDM model plus noise, as described by the provided covariance {{matrix}} once the intrinsic scatter term has been subtracted.

\section{Discussion and Conclusions}

In this work, we investigated the statistics of the Pantheon+ dataset and found that the $\chi^2$ of the best-fit $\Lambda$CDM model is significantly smaller than expected at the 3.9-$\sigma$ level.  This fact is also seen in  that the normalized residuals scatter less than would be expected from the Pantheon+ covariance matrix. Since there is no significant amount of kurtosis, it seems that the Pantheon+ covariance matrix is overestimated.  

If the covariance matrix is isotropically reduced by a factor of $0.93^2$, or if an intrinsic scatter term of $\delta^2=0.002$ is subtracted from the diagonal of the covariance matrix, then the expected $\chi^2$ value of 1580 is achieved.  Focusing on the latter case as it is more physically motivated, when we apply the iterative smoothing method to the real data with the subtracted covariance matrix, we find no significant departure from $\Lambda$CDM.  We should note that we expect that the overestimated errors of the SN distances in the Pantheon+ dataset should not significantly affect the process of parameter estimation and Bayesian model selection in conventional statistical analysis since such analysis is mainly sensitive to the $\Delta \chi^2$ and its distribution with respect to the best-fit points in confidence balls and not directly sensitive to the values of the best fit $\chi^2$ of models. This can be investigated in more detail in future works.

{One explanation for this overestimated covariance matrix comes from how statistical vs. systematic uncertainties are combined in the Pantheon+ analysis. This process was first proposed in ~\citep{2011ApJS..192....1C}.  To be specific, the statistical uncertainties are estimated such that, with only these statistical uncertainties, the reduced $\chi^2$ is 1~\citep{2022ApJ...938..110B}.  A covariance matrix for the systematic uncertainties is then added to the statistical one. It is this combined covariance matrix that yields an excessively small $\chi^2$ statistic. We have shown that this combined statistical and systematic covariance matrix does not represent the scatter of the residuals and thus may not be most suitable to calculate likelihoods for new cosmological analyses with these data. The Pantheon+ method of accounting for the intrinsic scatter before accounting for systematic errors explains why the ``statistical + systematic'' covariance is excessively large, and thus why the resulting $\chi^2$ value we find is small. The simplest fix would be reducing the intrinsic scatter term in the ``statistical+systematic'' covariance {{matrix}} by $\delta^2=0.002$, as we found in the analysis part of this paper.}

{%
However, this problem of the excessively large covariance matrix also raises the question of whether the systematic effects in supernova data are best accounted for by adding terms to the covariance matrix which is used to define a likelihood. Indeed, by the definition of the likelihood as the statistical distribution of the data given some model, systematic uncertainties do not affect the variance of the likelihood. 
Rather than account for these systematic effects by adding them to the covariance matrix, the systematic effects should be accounted for by parameterizing them and marginalizing over such parameters in a Bayesian manner.}

\textls[-15]{To be mathematically precise, consider the two approaches to accounting for systematics:}
\begin{equation} 
\mathcal{L}(\vect{\mu}_{\rm data}|\vect{\theta}_{c}) = \int d\vect{\psi}  \mathcal{L}(\vect{\mu}_{\rm data}|\vect{\theta}_{c}, \vect{\psi}) = \int d\vect{\psi}  \exp\left(- \frac{\vect{R}(\vect{\theta}_{c}, \vect{\psi})^\intercal \mat{C}_{\rm stat}^{-1}  \vect{R}(\vect{\theta}_{c}, \vect{\psi})}{2}\right), 
\end{equation}
where $\mathcal{L}$ is the likelihood, $\vect{\mu}_{\rm data}$ represents the SN lightcurve data reduced to distance moduli, $\vect{\theta}_c$ is a vector of cosmological parameters, $\vect{\psi}$ represents a systematic effect, $\vect{R}$ is the vector of residuals 
$\vect{\mu}_{\rm data} - \vect{\mu}_{\rm th}(\vect{\theta}_c, \vect{\psi})$, and $\mat{C}$ is the covariance matrix composed of a ``statistical'' term and a ``systematic'' term that depends on an uncertainty in the systematic effect, $\sigma_\psi$, as in Equations (6) and (7) of \cite{2022ApJ...938..110B}.
{{In contrast,} consider the alternative form:}
\begin{equation} 
\mathcal{L}(\vect{\mu}_{\rm data}|\vect{\theta}_{c}) = \exp\left(- \frac{\vect{R}(\vect{\theta}{c})^\intercal  \left(\mat{C}_{\rm stat} + \mat{C}_{\rm sys}(\sigma_\psi)\right)^{-1} \vect{R}(\vect{\theta}_{c})}{2}\right). 
\end{equation}
{These} two expressions are evidently not equivalent.


{The Pantheon+ team investigated some alternative ways to account for systematic effects in~\cite{Brout:2020bbg}, but these methods were found to lose information.
}



{An additional explanation for the smallness of the Pantheon+ $\chi^2$ is that selection effects, for instance a Malmquist bias, prevent the likelihood from being accurately described by a Gaussian~\cite{SupernovaCosmologyProject:2015zlj,2019ApJ...876...15H}.  In particular, the residuals at higher redshift would scatter less than expected from the intrinsic dispersion; only those SNe with the smallest distance moduli for a given redshift would be observed, not the full range.}

{We do not advocate that any one of these effects is the full answer; we merely point out the that this smallness of the $\chi^2$ effect exists within the Pantheon+ dataset, and  we write this brief paper to inform the larger cosmological community of it, and caution them when fitting models to it. }

\vspace{6pt} 


\authorcontributions{
R.E.K. wrote the code and performed the analysis. 
B.L. wrote the iterative smoothing code. 
A.S. oversaw the study.
All authors contributed to the writing of the paper. All authors have read and agreed to the published version of the manuscript.
}

\funding{%
A.S.  would like to acknowledge the support by the National Research Foundation of Korea (NRF-2021M3F7A1082053). 
B.L. acknowledges the support from the National Research Foundation of Korea (NRF-2022R1F1A1076338). 
A.S. and B.L. Acknowledge the support from {the National Research Foundation of Korea (RS-2023-00259422) 
and} the Korea Institute for Advanced Study (KIAS) grant funded by the government of Korea.}

 \dataavailability{{ }
The Pantheon+ data are publically available at \url{https://github.com/PantheonPlusSH0ES}.
 } 




\acknowledgments{%
We would like to thank Eric Linder and Alex Kim for useful discussions. We acknowledge Dillon Brout, Daniel Scolnic, and Adam Riess for useful discussions when finalizing \mbox{this draft.} 
}

\conflictsofinterest{The authors declare no conflicts of interest.} 





\begin{adjustwidth}{-\extralength}{0cm}
\setenotez{list-name=Note}
\printendnotes[custom] 


\reftitle{References}

\PublishersNote{}
\end{adjustwidth}
\end{document}